\documentclass[runningheads]{llncs}
\usepackage[T1]{fontenc}
\usepackage{graphicx}
\usepackage{times}
\usepackage{epsfig}
\usepackage{graphicx}
\usepackage{amsmath}
\usepackage{amssymb}
\usepackage{booktabs}
\usepackage{comment}
\usepackage{enumitem}
\usepackage{multirow}
\usepackage{hyperref}
\usepackage{makecell}
\usepackage{pifont}
\newcolumntype{M}[1]{>{\centering\arraybackslash}m{#1}}
\newcolumntype{L}[1]{>{\arraybackslash}m{#1}}
\usepackage{marvosym}

\newcommand{\name}{{\sc CS3}}

\begin{document}
	\title{CS3: Cascade SAM for Sperm Segmentation}

	% \author{anonymous}
	% %
	% \authorrunning{ }
	% \institute{anonymous}

	\author{Yi Shi\inst{1,2}\textsuperscript{({*})}, Xu-Peng Tian\inst{1,2}\textsuperscript{({*})}, Yun-Kai Wang\inst{1,2}, Tie-Yi Zhang\inst{1,2}, Bin Yao \inst{3,4}, \\Hui Wang\inst{3,4}, Yong Shao\inst{3,4}, Cen-Cen Wang\inst{3,4}, Rong Zeng\inst{3,4}, De-Chuan Zhan\inst{1,2}\textsuperscript{({\Letter})}}
	\authorrunning{Y Shi et al.}
	\institute{School of Artificial Intelligence, Nanjing University, Nanjing, China \and
		State Key Laboratory for Novel Software Technology, Nanjing University, Nanjing, China \and
		Jinling Hospital, Affiliated Hospital of Medical School, Nanjing University, Nanjing, China
		\and Jiangsu Provincial Medical Key Discipline Cultivation Unit, Nanjing, China \\
		\email{zhandc@nju.edu.cn}
	}

	\maketitle              % typeset the header of the contribution
	\begin{abstract}
		% Automated sperm morphology analysis is essential for efficiently evaluating male fertility but is hindered by suboptimal segmentation of sperm images. Current segmentation methods, such as the Segment Anything Model~(SAM), fail to address the challenge posed by overlapping sperm, which is common in clinical settings. Through preliminary experiments, we discover that modifying image properties, such as excluding sperm heads and easily segmentable regions, as well as enhancing the visibility of overlapping areas, significantly improve SAM's ability to segment complex sperm structures. Inspired by these insights, we introduce our Cascade SAM for Sperm Segmentation~({\name}), tailored to overcome the overlapping in sperm images. Through the cascade application of SAM, sperm heads, simple tails, and complex tails are segmented step by step. All these masks are then matched and spliced to generate complete masks. Collaborating with leading hospitals, we collect approximately 2,000 unlabeled sperm images for method refinement and obtain expert annotations on 240 additional images for model evaluation. Experimental results demonstrate superior performance of {\name} compared to existing methods. 
		
		Automated sperm morphology analysis plays a crucial role in the assessment of male fertility, yet its efficacy is often compromised by the challenges in accurately segmenting sperm images. Existing segmentation techniques, including the Segment Anything Model~(SAM), are notably inadequate in addressing the complex issue of sperm overlap—a frequent occurrence in clinical samples. Our exploratory studies reveal that modifying image characteristics by removing sperm heads and easily segmentable areas, alongside enhancing the visibility of overlapping regions, markedly enhances SAM's efficiency in segmenting intricate sperm structures. Motivated by these findings, we present the Cascade SAM for Sperm Segmentation ({\name}), an unsupervised approach specifically designed to tackle the issue of sperm overlap. This method employs a cascade application of SAM to segment sperm heads, simple tails, and complex tails in stages. Subsequently, these segmented masks are meticulously matched and joined to construct complete sperm masks. In collaboration with leading medical institutions, we have compiled a dataset comprising approximately 2,000 unlabeled sperm images to fine-tune our method, and secured expert annotations for an additional 240 images to facilitate comprehensive model assessment. Experimental results demonstrate superior performance of {\name} compared to existing methods. 
		
	\end{abstract}
	\let\thefootnote\relax\footnotetext{(*) means equal contribution to this work.}

	\section{Introduction}\label{sec:intro}

	Sperm morphology analysis is a crucial clinical technique for evaluating male fertility~\cite{gatimel2017sperm,eliasson2010semen,moruzzi1988quantification}. Currently, this analysis predominantly depends on the subjective evaluations of andrology specialists using microscopy, a process that is both time-consuming and labor-intensive, failing to satisfy the demands for large-scale, efficient assessments. Furthermore, the assessment of sperm morphology is highly subjective and lacks standardized criteria, presenting a significant challenge to reproductive medicine~\cite{sun2006there,van2015status,menkveld2011measurement}. Consequently, there is a pressing need for an automated system to assess sperm morphology in andrology clinical diagnostics, with the primary challenge being the precise segmentation of individual sperm within images. Sperm microscopic images frequently exhibit issues such as overlapping, indistinct boundaries, and various disruptive elements, which complicate the task for existing image segmentation models to accurately isolate sperm. This complexity further hampers the manual annotation of sperm image data, leading to a critical shortage of labeled sperm image datasets.
	
	Image segmentation technology has seen widespread application across various fields in recent years~\cite{patil2013medical,liu2019recent,ghosh2019understanding,minaee2021image,ma2024segment}, with its use in medical imaging becoming increasingly prevalent~\cite{wei2021deep,zeng2022semi,xie2022uncertainty,yu2020mammographic,DBLP:journals/iet-cvi/KheirkhahMS19}. In the specific context of sperm image segmentation, existing models can be categorized into three types. The first type~\cite{DBLP:journals/npl/LvYQLZZ22,shahzad2023sperm} focuses solely on segmenting the sperm head, neglecting the tail. The second type~\cite{DBLP:conf/hsi/FraczekKMLJM22,DBLP:journals/mbec/IlhanSSA20} is capable of segmenting only non-overlapping sperms, failing to address instances where sperms overlap. The third type~\cite{DBLP:journals/cbm/LewandowskaWMLJ23,DBLP:journals/iet-cvi/KheirkhahMS19} treats overlapping sperms as a single entity, without distinguishing between individual sperms. These approaches all rely on labeled data for training, yet the acute scarcity of labeled sperm image datasets has led to a simplification in their segmentation targets. However, sperm overlap is a common occurrence in clinical settings, rendering these methods inadequate for practical clinical applications. The ability to effectively segment overlapping sperms, generating independent and complete masks for each sperm without relying on labeled training data, is crucial for automated sperm morphology analysis.
	
	The sperm image segmentation task is essentially an instance segmentation task where the goal is to acquire distinct masks for each sperm without necessarily differentiating their semantic attributes. Recent advancements, such as CutLER~\cite{wang2023cut} and U2Seg~\cite{Niu2023U2Seg}, have demonstrated promising results in unsupervised instance segmentation within common scenarios. However, these techniques often fall short in more specialized medical contexts, such as distinguishing sperms. The Segment Anything Model (SAM)~\cite{Alexander2023SAM} represents a significant leap forward in image segmentation technology. Operating in everything mode, SAM can segment images it has never seen before without any annotations or prompts. This capability has led to successful applications across various domains~\cite{liu2023matcher,zou2024segment,archit2023segment,cheng2023sam,wong2023scribbleprompt,wang2023cut}. Nonetheless, our experiments indicate that SAM tends to segment only those tails that are distinct and non-overlapping, neglecting or inaccurately grouping overlapping tails without further segmentation. This limitation highlights the need for tailored approaches to address the unique challenges of sperm image segmentation.
	
	To tackle the challenges identified in sperm image segmentation, particularly with the SAM in everything mode, we begin with preliminary experiments that yield three pivotal insights. Firstly, We find that SAM will give priority to segmentation by color, and will mainly consider geometric features when there is no obvious difference in color. Secondly, we discover that excluding easily segmentable regions from the original image prompts SAM to target more complex areas. Lastly, we find that enlarging and thickening the lines of overlapping sperm tails render these previously indistinct areas separable. Leveraging these insights, we introduce our method {\name}, a cascade SAM algorithm for the end-to-end segmentation of sperm images. This unsupervised approach effectively mitigates the issues posed by overlapping sperms, enabling precise differentiation and segmentation of individual sperms, including both heads and tails, to generate independent and complete masks for each. {\name} begins with an initial segmentation using SAM on pre-processed images, followed by the application of color filters to isolate sperm head masks. These masks are then saved and removed from the image, leaving an image with only sperm tails. Subsequently, the cascade process focuses on sperm tails, each time isolating and preserving masks that adhere to the morphological criteria of individual tails, thus allowing for a progressive segmentation from simpler to more complex forms until no further changes are observed between two successive rounds. At this stage, while most tails are independently segmented, a few remain intertwined. For these, we apply an enlargement and line-thickening technique before subjecting them to further segmentation with SAM, with the segmented results then resized to their original dimensions. Finally, {\name} matches the obtained head and tail masks based on criteria such as distance and angle to assemble complete and independent sperm masks. To our knowledge, this cascade approach using SAM for segmenting intersecting elongated structures, such as those found in sperm tails, is novel. Moreover, {\name} offers a fresh perspective for segmenting similar structures in other domains, such as vascular and neural imaging.
	
	In collaboration with several leading hospitals, we collect approximately 2,000 unlabeled sperm images to refine the parameters of our proposed method. Additionally, we engage relevant experts to meticulously annotate 240 sperm images, serving as a benchmark for model evaluation. The empirical evidence suggests that our {\name} surpasses previous methodologies in segmentation efficacy, particularly in segmenting overlapping sperms. This advancement paves the way for leveraging artificial intelligence to fully automate sperm morphology analysis.
	
	Our contributions are articulated as follows:
	(1) We identify three limitations of SAM in sperm segmentation and provide actionable solutions.
	(2) We develop an unsupervised method named {\name} for sperm image segmentation, employing cascade applications of SAM.
	(3) Experiments demonstrate that {\name} outperforms existing methods in segmentation accuracy, especially in resolving overlapping sperm instances, marking a significant advancement in the field.
	
	\section{Method}\label{sec:method}
	
	In this section, we present a detailed exposition of our {\name}, a novel algorithm designed for the unsupervised segmentation of sperm images using cascade applications of SAM. Initially, we highlight three critical experimental observations identified during the application of SAM in everything mode for the task of sperm image segmentation. These observations serve as foundational insights that inspire the development and comprehensive delineation of the {\name} process.

	\subsection{Preliminary Investigations into SAM}\label{sec:SAM}
\begin{figure}[t]
		\centering
		\begin{minipage}[c]{\linewidth}
			\includegraphics[width=\linewidth]{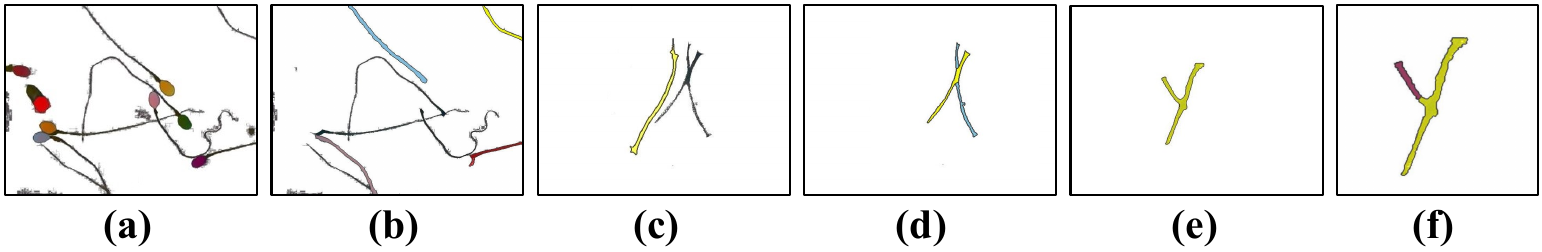}\\
		\end{minipage}
		\caption{Preliminary Investigations into SAM on pre-processed sperm images. (a)-(b) After removing sperm heads, SAM begins to segment remaining parts from a geometric perspective. (c)-(d) After removing simple parts, SAM begins to focus on complex areas. (e)-(f) After enlarging and bolding, overlapping parts that are difficult to split become separatable.
		}\label{fig:1}
	\end{figure}

	In our exploratory studies, we employ SAM in everything mode to perform segmentation on clinical sperm images. Raw images are initially pre-processed for better performance of SAM. Our results show that SAM effectively segments sperm heads but struggles with tail segmentation. We speculate that SAM will give priority to segmentation based on color, so that the segmentation effect on sperm heads with obviously different colors is better, which is also consistent with the findings in~\cite{ji2023segment,archit2023segment,deng2023segment,ma2024segment}. When we try to remove sperm heads from the image so that the remaining parts have similar colors, SAM intends to divide it from a geometric perspective. At this time, SAM can segment some simple sperm tails, as shown in Figure~\ref{fig:1}(a)-(b). Notably, SAM's capability is confined to the identification of distinct tails, showing limitations in recognizing overlapping sperm tails or in delineating them into separate entities. To enhance SAM's segmentation proficiency for sperm imagery, modifications are attempted on the tail images. Figure~\ref{fig:1}(c)-(d) demonstrates that omitting the readily segmentable tails from the imagery prompts SAM to shift focus towards the intricate tails, thereby initiating their segmentation. This discovery inspires us to design a cascade structure to segment the entire image step by step. Besides, Figure~\ref{fig:1}(e)-(f) shows that in some cases SAM faces challenges in segmenting overlapping slender tails into distinct entities. We find that enlarging these regions and bolding the slender structures enable SAM to recognize and further partition these areas. All these experimental observations are instrumental in the conceptualization and formulation of the {\name} algorithm.\\
	
	\subsection{Cascade SAM for Sperm Segmentation~({\name})}\label{sec:CS3}
\begin{figure}[t]
		\centering
		\begin{minipage}[c]{\linewidth}
			\includegraphics[width=\linewidth]{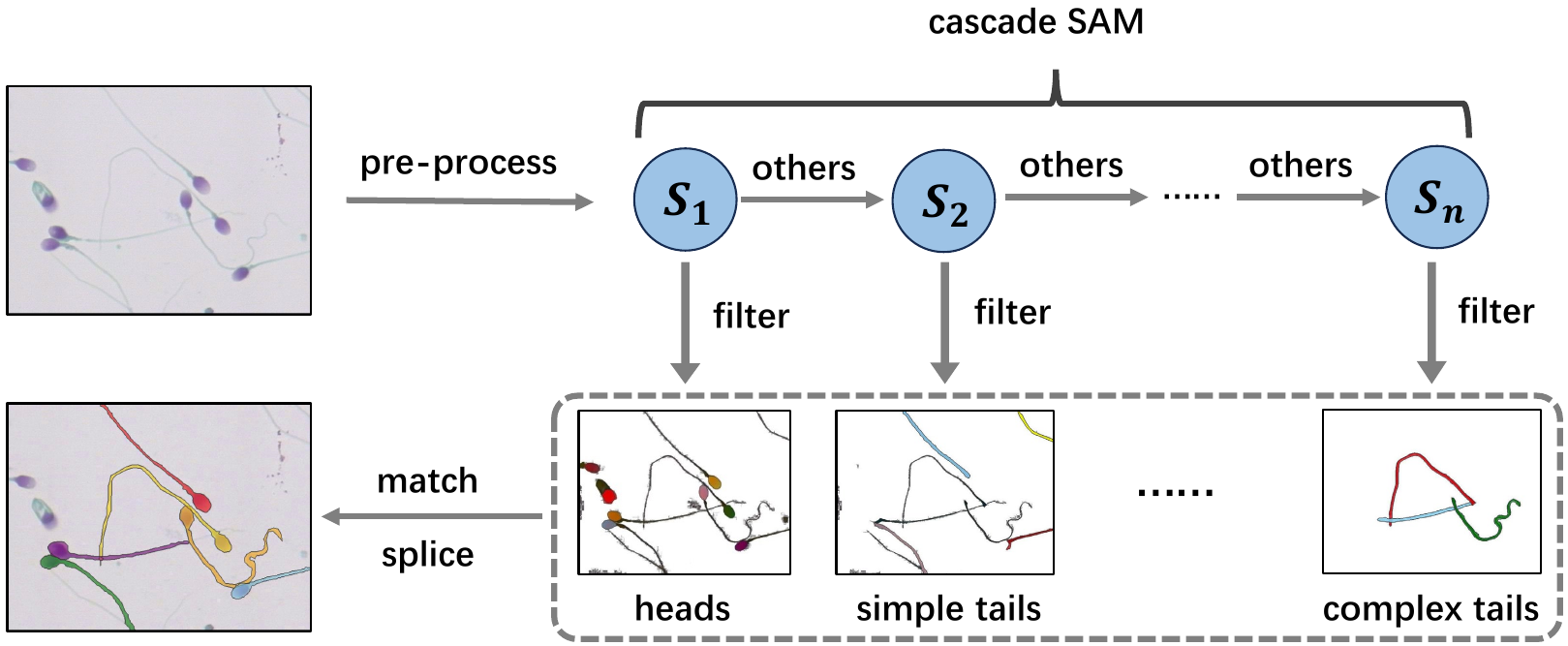}\\
		\end{minipage}
		\caption{The architecture of our method {\name}. Through the cascade application of SAM, sperm heads, simple tails, and complex tails are segmented step by step. All masks are matched and spliced to generate complete masks. $S_{1}, S_{2}, \cdots, S_{n}$ donotes a series of SAM in everything mode.}\label{fig:2}
	\end{figure}

	The architecture of our {\name} is shown in Figure~\ref{fig:2}. Initially, a series of preprocessing steps are applied to the raw images to enhance quality for segmentation. These steps include adjustments to brightness, contrast, and saturation, along with background whitening, aimed at reducing noise and emphasizing the primary features of the sperm. Then, {\name} uses a sequence of SAMs, denoted as $S_{1}, S_{2}, \ldots, S_{n}$, for stepwise segmentation of pre-processed images.
	Inspired by our first observation, $S_{1}$ is used to segment sperm heads, simplifying the remaining parts' colors for easier geometric division by subsequent SAM applications. By intersecting the obtained masks from $S_{1}$ with the purple regions of the raw image and applying a threshold filter based on the intersection's proportion, we efficiently isolate all head masks. These masks are kept and then eliminated from the original image, resulting in an image that only contains sperm tails.
	
	The process from $S_{2} $ to $S_{n}$ takes full advantage of our second preliminary observation. After each round of SAM, the newly acquired masks undergo a filtration process. This step involves preserving and discarding single tail masks, thereby focusing on overlapping tails and those not yet detected by SAM. To accurately identify single tail masks, obtained masks are first skeletonized into one-pixel-wide lines. These lines are then subjected to two critical filtration criteria: (1) the presence of a single connected segment, and (2) the line terminating in {\em exactly} two endpoints. The first criterion ensures the exclusion of masks representing multiple aggregated tails, while the second confirms the mask outlines a solitary tail. 
	% The determination of endpoints within a line utilizes a $3 \times 3$ convolution kernel, identifying endpoints as pixels connected to only one adjacent pixel. 
	Masks conforming to these specifications are deemed single tail masks, saved, and then removed from the original image. Following this extraction, the image undergoes a denoising process to enhance clarity and reduce potential segmentation errors. This cascade process persists until SAM's segmentation outputs remain consistent across two successive rounds. 
	
	The cascade SAM methodology delineated herein proves effective in resolving the majority of instances involving overlapping sperm. However, a marginal subset of these overlaps presents a notable challenge, resisting separation through cascade processing. To address these particularly resilient cases, {\name} uses the enlargement and bold method inspired by our third preliminary observation to solve the problem in the last round of SAM. Specifically, we first use the following two rules to filter out those parts that still overlap: (1) the presence of a single connected segment, and (2) the line terminating in {\em more than} two endpoints. Each selected mask is then emphasized by enlarging its presence within the image and augmenting the outline's width. This process involves isolating the mask as the foreground, cropping the background to enhance the foreground's relative size, and applying edge detection to delineate and thicken the outline. Subsequent smoothing of the enlarged outline mitigates any resultant jaggedness. These tailored steps facilitate a further round of SAM segmentation to these obstinate overlapping areas, with the segmented masks thereafter resized and repositioned to their original context within the image.
	
	\begin{figure}[t]
		\centering
		\begin{minipage}[c]{\linewidth}
			\includegraphics[width=\linewidth]{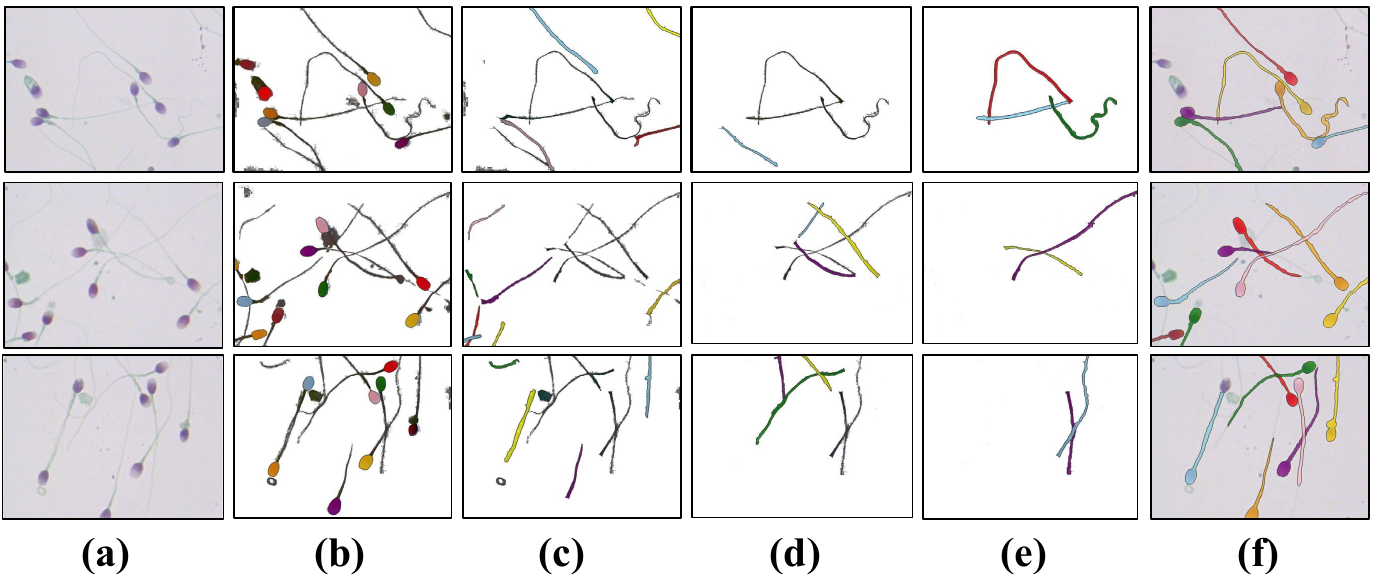}\\
		\end{minipage}
		\caption{Segmentation results of {\name}. (a) The raw images. (b)-(e) The results of cascade SAM gradually segmenting different parts of sperm from simple to complex. (f) The results after mask matching and splicing. More results can be found in the appendix.
		}\label{fig:3}
	\end{figure}

	The culmination of this process entails the matching and splicing of the corresponding head and tail masks to construct comprehensive sperm masks. To achieve this, an elliptical fitting technique is applied to delineate each sperm head, leveraging the ellipse's major axis to closely approximate the head's geometry. Tail structures are represented by their skeletonized lines. Subsequently, the endpoints of these skeletonized tails and the major axes of the heads are identified, with particular emphasis on the terminal slopes at these endpoints. Our matching process is grounded in the hypothesis that a minimal distance between endpoints and a similarity in the terminal slopes indicate a higher likelihood of the endpoints belonging to the same sperm entity. Therefore, we establish a distance threshold $\lambda_{dis}$ and an angle threshold $\lambda_{angle}$ for filtering. In cases where multiple endpoints meet these criteria for a single anchor endpoint, preference is given to the one closest in angle. Successful matching yields the assembly of complete sperm masks, effectively bridging the head and tail components. All results in the process of CS3 are shown in Figure~\ref{fig:3}.

	\section{Experiment}\label{sec:exp}
	
	\noindent\textbf{Dataset.}
	In a collaborative effort with leading hospitals, we collect a dataset of approximately 2,000 unlabeled sperm images directly from clinical practice, vital for our model's parameter optimization. In addition, relevant andrology experts also annotate an additional 240 sperm images, and the annotation results of each image are checked and approved by at least three senior experts for model evaluation. These original images are all 720*540 pixels in size. 
	% In order to facilitate the effectiveness of the segmentation algorithms, we pre-processed the images before running {\name} and all comparison methods, including increasing the resolution, unifying the brightness value, adjusting the contrast and saturation of the image, sharpening the edges, and whitening the background, etc. 
	For those comparing methods that require labeled samples for training, we divide all 240 existing labeled images according to 3:1, that is, 180 images are used for their training, and 60 images for their performance evaluation.\\
	
	\noindent\textbf{Evaluation metric.}
	Drawing on previous studies~\cite{DBLP:conf/hsi/FraczekKMLJM22,DBLP:journals/mbec/IlhanSSA20,DBLP:journals/npl/LvYQLZZ22}, this research adopts two prevalent evaluation metrics in the domain of instance segmentation: the mean Intersection over Union~(mIOU) and the mean Dice coefficient~(mDice). Here, ``mean'' signifies the computation of average metric values across all instances. mIOU and mDice assess a model's accuracy by measuring the extent of overlap between the actual areas (denoted as $A_{i}, i=1,\cdots,N$) and the predicted areas (denoted as $B_{i}, i=1,\cdots,N$). A greater overlap corresponds to a higher metric value, indicating superior model performance. The specific definitions are as follows:
	
	\begin{equation*}
		\operatorname{mIoU}=\frac{1}{N} \sum_{i=1}^N\left(\frac{\left|A_i \cap B_i\right|}{\left|A_i \cup B_i\right|}\right), \; \text { mDice }=\frac{1}{N} \sum_{i=1}^N\left(\frac{2 \times\left|A_i \cap B_i\right|}{\left|A_i\right|+\left|B_i\right|}\right)\;.
	\end{equation*}
	Before calculating mIOU and mDice, we ascertain the optimal pairing relationship by comparing the IOU ratios between each instance in the Ground Truth and the predicted results. It is important to note that the segmentation outcomes of Ground Truth and some methods may not correspond on a one-to-one basis, potentially resulting in disparate quantities. In such cases, we employ a filtering criterion predicated on the logic that pairs with an optimal pairing IoU ratio are selected for calculation.\\
	
	\noindent\textbf{Comparison methods.}
	This study employs two types of principal methodologies for comparison. The first type centers on supervised learning strategies derived from U-net~\cite{ronneberger2015u}, extensively applied on sperm segmentation, such as IUHS~\cite{DBLP:journals/npl/LvYQLZZ22} and CN2UA~\cite{marin2021impact}. Owing to the unavailability of source codes, this research undertake independent replication of IUHS and CN2UA, guided by the detailed descriptions provided in these papers. The second type encompasses unsupervised instance segmentation, such as SAM~\cite{Alexander2023SAM}, U2Seg~\cite{Niu2023U2Seg}, and CutLER~\cite{wang2023cut}. For these approaches, the deployment of models are facilitated by utilizing source codes provided by the original authors.\\
	
	\noindent\textbf{Implementation details.} 
	Through statistical analysis of unlabeled images, we identify the HSV~(Hue, Saturation, Value) range for detecting purple regions as $(100.0-180.0, 20.0-255.0, 20.0-255.0)$. This range is crucial for the filtering process in generating head masks. Furthermore, we employ a $3 \times 3$ convolution kernel to determine line endpoints, defining them as pixels connected to only one neighboring pixel. The distance threshold $\lambda_{dis}$ and the angle threshold $\lambda_{angle}$ are set to $20$ pixels and $60$ degrees, respectively. The enlarging and bolding as well as the corresponding restoration steps are all completed automatically by Python code without manual intervention. Our analysis indicates that images in our test set require an average of $4.8$ rounds of SAM for optimal processing, with some needing up to $7$ rounds. Codes of {\name} are available at https://github.com/shiy19/CS3.\\
	
	\begin{table}[t]
		\centering
		\caption{mIOU and mDice of different methods on our collected sperm image dataset. The best results are in bold.}
		\tabcolsep 8pt
		\begin{tabular}{c|ccccccc}
			\addlinespace
			\toprule
			Method & U-net & IUHS & CN2UA & SAM & U2Seg & CutLER & CS3\\
			\toprule
			mIOU  & 49.21 & 47.56  & 19.26 & 34.17 & 16.44 &31.21 & {\bf 72.50} \\
			mDice & 62.17 & 58.46  & 32.94 & 46.75 & 28.86 & 44.17& {\bf 80.26}\\ 
			\bottomrule
		\end{tabular}\label{table:exp}
		% \vspace{-3mm}
	\end{table}
	
\begin{figure}[t]
		\centering
		\begin{minipage}[c]{\linewidth}
			\includegraphics[width=\linewidth]{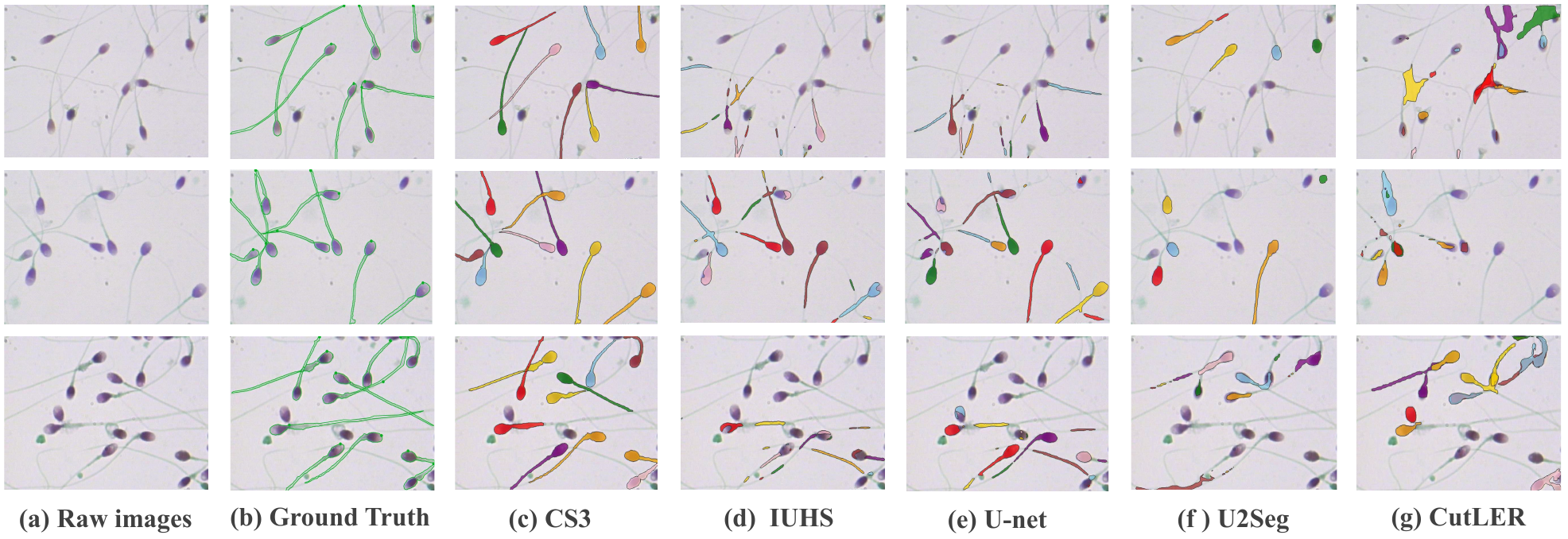}\\
		\end{minipage}
		\caption{Segmentation results of different methods. More results can be found in the appendix.
		}\label{fig:4}
	\end{figure}

	\noindent\textbf{Performance evaluation and visualization.} 
	The experimental findings on our sperm dataset are illustrated in Table \ref{table:exp}, where both the comparative methods and our {\name} are applied to pre-processed images. Supervised techniques such as U-Net, IUHS, and CN2UA necessitate extensive labeled datasets for training, rendering them less effective on sperm datasets that lack substantial labeled data. These methods often oversimplify the complex task of sperm segmentation, leading to their ineffectiveness on our real-world clinical sperm dataset. On the other hand, unsupervised instance segmentation techniques like SAM, U2Seg, and CutLER depend on recognizing texture and structural patterns within the images. However, sperm images significantly diverge from conventional natural images due to their high contrast, scant texture, and the interference caused by dye blocks. These distinct characteristics hinder the learning process from such images. While some methods utilize advanced data augmentation strategies to introduce variability, these attempts can inadvertently remove essential features of sperm imagery, such as tail position and morphology, thereby degrading the model's performance. Additionally, the common occurrence of numerous elongated tails and their uneven distribution complicates the task of accurately characterizing these structures. The comparative methods generally struggle with identifying and processing these complex, overlapping structures. In contrast, our {\name} adeptly segments sperm images by progressively handling simpler to more challenging regions through the cascade application of SAM, particularly excelling in segmenting overlapping parts. By effectively assembling the head and tail components, {\name} achieves independent and complete segmentation of individual sperm without relying on labeled data for training. Therefore, our {\name} results in superior performance across various evaluated metrics. Visual results achieved by different methods are displayed in Figure~\ref{fig:4}.
	
	\section{Conclusion}\label{sec:con}
	We experimentally explore the difficulties of SAM in sperm image segmentation and propose effective solutions. Based on these experimental observations, we develop an unsupervised algorithm named {\name} for the automated segmentation of sperm morphology images, specifically addressing the challenge of overlapping. Through the cascade application of SAM, {\name} can gradually segment different parts of the sperm image from simple to difficult, and then combine these segmentation results into complete sperm masks. Our experimental results on a clinical sperm dataset collected in-house demonstrate superior segmentation performance of {\name} compared to previous methods. Future work will focus on researching automated sperm morphological classification methods based on the segmentation results from {\name}. 
	\\~\\
	\textbf{Limitations.}  
	{\name} faces challenges in processing images with excessively complex tail overlaps, such as scenarios where more than ten sperm are entangled, a situation that even human experts find difficult to discern. Therefore, adequate dilution of semen samples is still necessary in clinical practice to ensure manageable analysis conditions.
	\\~\\
	\textbf{Acknowledgments.} This research was supported by NSFC (61921006), Clinical Technology Fundation of Jinling Hospital (22LCZLXJS68), and Collaborative Innovation Center of Novel Software Technology and Industrialization.

	\bibliographystyle{splncs04}
	\bibliography{mybib}
	
\end{document}